\documentclass{JHEP3}
\relax

\def\bes{\begin{eqnarray}}
 \def\ees{\end{eqnarray}}
\def\be{\begin{equation}}
\def\ee{\end{equation}}
\def\bs{\begin{subequations}}
\def\es{\end{subequations}}
\newcommand{\een}{\end{subequations}}
\newcommand{\ben}{\begin{subequations}}
\newcommand{\beq}{\begin{eqalignno}}
\newcommand{\eeq}{\end{eqalignno}}

 \def\lx{\lambda}
 
 \def\kt{{\tilde{k}}}

\def \gta {\mathrel{\vcenter
     {\hbox{$>$}\nointerlineskip\hbox{$\sim$}}}}

\title{The temperature and entropy of CFT on time-dependent backgrounds}

\author{N. Tetradis$^{a,b}$
\\
$^a$Department of Physics, University of Athens, Zographou 157 84, Greece 
\\
$^b$Department of Physics, Stanford University, Stanford, CA 94305, USA
\\
{\tt ntetrad@phys.uoa.gr}}

\preprint{}

\abstract{
We express the AdS-Schwarzschild black-hole configuration in coordinates
such that the boundary metric is of the FLRW type. We review how this construction 
can be used in order to calculate the
stress-energy tensor of the dual CFT on the FLRW background. We 
deduce the temperature and entropy of the CFT, 
which are related to the temperature and entropy
of the black hole. We find that the entropy 
is proportional to the area of an apparent horizon, different
from the black-hole event horizon. 
For a dS boundary we reproduce
correctly the intrinsic temperature of dS space.}


\begin{document}

\section{Introduction}

The correspondence between supergravity on anti-de Sitter (AdS) space and 
boundary conformal field theory (CFT) 
\cite{adscft} permits the derivation of many properties of CFT through the 
study of classical gravity solutions. In this work our interest lies in the 
thermodynamics of strongly coupled gauge theories, such as the
$\mathcal{N}=4$ super Yang-Mills theory with $SU(N)$ gauge symmetry in the large-$N$
limit, 
in four dimensions. The thermodynamics can be deduced from 
the properties of a Schwarzschild black hole in
five-dimensional AdS space \cite{witten}. For a Minkowski boundary the 
CFT entropy is proportional to the volume of the black-hole event horizon, 
in agreement with the standard notions of black-hole entropy \cite{bh}
and holography \cite{holograph}. 

An important application of the AdS/CFT correspondence concerns the holographic 
description of hydrodynamic properties of field theories. Of particular interest are
the properties of the quark-gluon plasma (QGP) produced in heavy-ion collisions. 
The system can be modelled through the so-called Bjorken flow \cite{bjorken}. 
The construction of the space-time dual to this flow \cite{janik,kinoshita}
has led to the derivation of properties of the plasma, such as the ratio of 
shear viscosity to entropy density \cite{son}. The holographic description 
of more general hydronamic regimes of strongly coupled gauge theories has
also been studied \cite{hubeny1}. 
The construction of gravity solutions dual to the Bjorken flow or its
generalizations relies on a perturbative expansion. 
Verifying the regularity of a solution, or
extracting physical information, is a difficult task. 
In general, one finds that the gravity duals are time-dependent.
They also possess event and apparent horizons that can be distinct. It is not obvious
which horizon is relevant for properties of the gauge theory, such as its
entropy \cite{hubeny2}. 

The above difficulties have motivated us to study the properties of a CFT on 
a time-dependent background, for which the gravity dual is exactly known. 
The theory is the standard $\mathcal{N}=4$ super Yang-Mills theory at non-zero 
temperature on a Friedmann-Lemaitre-Robertson-Walker (FLRW) background.
The gravity dual is the AdS-Schwazschild geometry, but in a parametrization that
sets the boundary metric in the FLRW form. The appropriate form of the metric was
given in \cite{ast}, where the stress-energy tensor of the dual CFT was
also derived. The boundary metric can be made dynamical by including a boundary
Einstein term and imposing mixed boundary conditions along the lines of 
\cite{marolf}. This construction leads to the effective 
Friedmann equation governing the cosmological evolution, in which the quantum
properties of the CFT, such as the conformal anomaly and the Casimir energy, are
reflected \cite{ast}. 

In this work we derive the temperature and entropy of the CFT on a FLRW 
background. The temperature is deduced by requiring the smoothness of the 
five-dimensional metric. When the temperature and stress-energy tensor of the CFT
are known, the entropy can be computed. It is not neccesary to assume that the
entropy is proportional
to the surface of the black-hole horizon. Moreover, between the
distinct event and apparent horizons, we can identify
the one that is relevant for the CFT on a time-dependent background.

The temperature and entropy 
can display non-trivial behavior as a function of time. 
This may appear surprising, as the naive expectation is 
that the temperature would be simply redshifted by one power of the
scale factor $a(\tau)$ relative to
its value in the static case. The entropy density would fall proportionally to
$a^{-3}$, with the total entropy being conserved. However, the FLRW background
introduces a new scale in the problem, i.e. the expansion rate $\dot{a}/a$. 
When this becomes siginificant relative to the energy density of the CFT, the
temperature and entropy receive corrections. For example, 
in the particular case of exponential 
expansion (de Sitter space), the temperature is proportional to the constant $H=\dot{a}/a$. 
The main purpose of this work is to 
calculate these corrections and discuss their physical interpretation.

In the following warm-up section we discuss the case of a static boundary and
derive the temperature and entropy of the CFT.
In section 3 we summarize the
formalism that permits the study of a FLRW boundary, following \cite{ast}.
The basic formalism was derived in \cite{ast}, but we include it here for completeness.
We also discuss the event and apparent black-hole horizons.
In section 4 we derive the black-hole temperature, which we identify with the
temperature of the CFT. In section 5 we derive and discuss the CFT entropy
for a number of possible types of evolution
of the boundary.
The final section includes the discussion and interpretation of our results.
 
We emphasize that our analysis focuses on an expanding boundary with $\dot{a}\geq 0$.
Our general framework allows for both signs of $\dot{a}$. However, the discussion
of the geometry and its horizons, as well as the specific forms of $a(\tau)$ that we
consider assume $\dot{a}\geq 0$. This assumption parallels the Bjorken-flow geometry,
which is the motivation for our work. 

The questions we are 
addressing have also been discussed in \cite{kajantie}.
Our investigation has been carried out independently and applies to 
a general FLRW cosmology. A major difference between the two approaches 
lies in the
calculation of the temperature. We compute the CFT temperature from the 
five-dimensional metric. Our result incorporates corrections that become
important for large curvature. These corrections are reflected in the
CFT entropy, and lead to interesting novel behavior.

\section{Static boundary}

We consider an AdS-Schwarzschild black hole in five dimensions, 
which is a solution of the Einstein field equations with a 
negative cosmological constant 
($\Lambda_5 = -6/l^2$),
\be\label{eqein} R_{AB} = \frac{2}{l^2} g_{AB} \ , \ee
where $A,B = 0,1,2,3,4$.
The metric can be written in Schwarzschild coordinates as
\be\label{eqmetric} ds^2 = -f(r) dt^2 + \frac{dr^2}{f(r)} 
+ r^2 d\Omega_k^2 \  ,\ \ \ \ \ \ f(r) = r^2+k - \frac{\mu}{r^2} \ , \ee
where $k=+1,0,1$
for spherical, flat and hyperbolic horizons, respectively. We have 
set $l =1$ for simplicity.
The Hawking temperature can be computed through the elimination of the conical singularity in the 
Euclidean version of the metric \cite{page}. The energy (or mass) of the black hole can be computed through 
the temperature dependence of the gravitational action on the AdS-Schwarzschild background, after 
subtraction of the action of pure AdS space with compactified time \cite{witten,page}.
The temperature and energy are, respectively,
\be\label{eqTM} T = \frac{2r_e^2 + k}{2\pi r_e} \ , \ \ \ \ \ \ \ 
E = \frac{3V_k}{16\pi G_5} r_e^2 (r_e^2 + k) =\frac{3\mu V_k}{16\pi G_5}\ ,\ee
where $V_k$ is the volume of the three-dimensional space $\Sigma_k$ spanned by 
$\Omega_k$, $r_e$ is the radius of the event horizon, satisfying $f(r_e)=0$, and $G_5$ 
is the five-dimensional Newton's constant.

The case of a spherical horizon with $k=1$ is important for the 
interpretation of the
results of the following sections. 
For increasing $\mu \gg 1$ the temperature $T$ increases as
$\mu^{1/4}/\pi$. For decreasing $\mu \ll 1$, $T$ again increases
as $1/(2\pi\,\mu^{1/2})$. There are no black holes with $T$ 
below a minimal value $\sqrt{2}/\pi$. For larger $T$,
there are two black-hole solutions with the same temperature and different masses.
The one with lower mass is known to be unstable \cite{page}.
The solution with the larger mass is dual to the high-temperature
deconfined phase of the gauge theory \cite{witten}.

In order to discuss the AdS/CFT correspondence, we must express the metric in 
terms of Fefferman-Graham coordinates \cite{fg}. We define $z$ through
\be \frac{dz}{z} = -\frac{dr}{\sqrt{f(r)}}, \ee
which gives (with an appropriate integration constant)
\be \label{z4r} z^4 = \frac{16}{k^2+4\mu} \ \frac{r^2 + \frac{k}{2} - r\sqrt{f(r)}}{r^2 
+ \frac{k}{2}+ r\sqrt{f(r)}} \ . \ee
This equation may be inverted to give
\be\label{eq4a} r^2 = \frac{\alpha + \beta z^2 + \gamma z^4}{z^2}, \ee
where
\be\label{eq4x}
\alpha = 1 \ , \ \ \ \ \ \ \ \ \ \
\beta = - \frac{k}{2} \ , \ \ \ \ \ \ \ \ \ \ 
\gamma = \frac{k^2+4\mu}{16}\ . \ee
The metric (\ref{eqmetric}) reads
\be\label{eqmetric1} ds^2 = \frac{1}{z^2} 
\left[ dz^2 - \frac{\left( 1- \gamma z^4 \right)^2}{1+\beta z^2 + \gamma z^4} dt^2 
+ \left( 1+ \beta z^2 +\gamma z^4 \right) d\Omega_k^2 \right] \ . \ee

The coordinates $(\tau,z)$ are closely related to the isotropic coordinates that are 
often employed for the study of the Schwarzschild geometry.
The isotropic coordinates do not span the full space. 
They cover the two regions of the Kruskal-Szekeres plane that are located
outside the horizons.   
The same happens for the coordinates $(\tau,z)$ in the case of the
AdS-Schwazschild geometry. As 
can be deduced easily from eq. (\ref{z4r}), the coordinate $z$ corresponds to
values of $r$ that satisfy $f(r)\geq 0$. On the other hand, eq. (\ref{eq4a})
indicates that the region $(r_e,\infty)$ of $r$ is covered twice by the
coordinate $z$ taking values in $(0,\infty)$. For fixed coordinate time $t$, the metric describes
the Einstein-Rosen bridge connecting two asymptotically flat regions \cite{mtw}. In the following we shall
generalize this construction in a dynamical setting.

It is instructive to recall how the Hawking temperature $T$ 
of the black hole \cite{page} can be
determined from the metric (\ref{eqmetric1}). The partition function of a static 
thermalized
system can be obtained if we switch to Euclidean time and assume that the time direction
is periodic
with period $1/T$. For arbitrary $T$, the resulting   
metric possesses a conical singularity at the location of the horizon
$z_e= \gamma^{-1/4}$. This can be eliminated for a specific value of $T$, which
is identified with the natural temperature of the black hole. 
Expanding $z$ around $z_e$, we find that the conical
singularity disappears for $T$ given by 
\be\label{htz}
T=\frac{1}{\sqrt{2}\, \pi } \left(
\frac{k^2+4\mu}{\left(k^2+4\mu \right)^{1/2}-k}
\right)^{1/2},
\ee
in agreement with the first of eqs. (\ref{eqTM}).

The energy and entropy of the black hole can be related 
to those of the dual CFT on the AdS boundary. For a metric in the form
\be\label{eq2} ds^2 = \frac{1}{z^2} \left[ dz^2 + g_{\mu\nu} dx^\mu dx^\nu \right], \ee
where
\be g_{\mu\nu} = g_{\mu\nu}^{(0)} + z^2 g_{\mu\nu}^{(2)} 
+ z^4 g_{\mu\nu}^{(4)} + \dots \ee
the stress-energy tensor of the CFT is \cite{Skenderis}
\begin{eqnarray}
 T_{\mu\nu}^{(CFT)}  =
\frac{1}{4\pi G_5} 
\Biggl\{ g^{(4)} 
&-& \frac{1}{2} g^{(2)}  g^{(2)}
+ \frac{1}{4} \rm{Tr} \left[ g^{(2)}\right]\, g^{(2)} \Biggr.
\nonumber \\
&-& 
\Biggl.
\frac{1}{8} \left( \left(\rm{Tr} \left[g^{(2)}\right] \right)^2  
- \rm{Tr} \left[g^{(2)} g^{(2)}\right]  \right) g^{(0)} 
\Biggr\}_{\mu\nu} \ , 
\label{eq3a} \end{eqnarray}
with indices lowered and raised by $g^{(0)}$ and its inverse.
Applying this general expression to our metric (\ref{eqmetric1}), 
we obtain the energy density and pressure, respectively,
\be\label{eq3}  T_{tt}^{(CFT)}  = 3 T_{ii}^{(CFT)}  = 
\frac{3\left(k^2+4\mu \right)}{64\pi G_5} \ ,  \ee
on a four-dimensional space with metric
\be\label{eqmetric0} ds_0^2 = g_{\mu\nu}^{(0)} dx^\mu dx^\nu = -dt^2 + d\Omega_k^2 \ .\ee
Notice that the total energy $E= T_{tt}^{(CFT)}  V_k$ is larger 
than the mass of the black hole (eq.~(\ref{eqTM})) by a constant (Casimir energy) in the 
case of a curved horizon ($k\ne 0$). 
The two quantities agree for flat horizons ($k=0$). 

In order to compute the entropy of the CFT
we start from the thermal energy $E$, which can be determined from the partition function
$Z$ through the relation $E=-\partial (\ln Z)/\partial(1/T)$.
We need not compute $Z$, as we have already determined the energy. 
The entropy is given by $S=E/T+\ln Z$, where we have omitted the Casimir contribution
to the energy. The temperature is a function of $\mu$. 
Differentiating with respect to 
$\mu$ we obtain $dS/d\mu=(1/T)dE/d\mu$. A simple integration gives
\be\label{entropy}
S=\frac{V_k}{4G_5} r_e^3. 
\ee
As expected, the entropy is proportional to the surface of the event horizon.

In an alternative, more intuitive derivation we consider an infinitesimal 
variation of the 
parameter $\mu$ of the metric (\ref{eqmetric1}).
The volume $V_k$ of the boundary is not affected. The variation of $\mu$ generates
a variation of the internal energy $E$ of the system that can be attributed to
a change of its entropy $S$. Assuming that the process takes place sufficiently slowly,
the variations of 
$E$ and $S$ obey $dE=TdS$. The integration with respect to $\mu$ gives
eq. (\ref{entropy}).
It must be pointed out that the above determination of the entropy takes into
account only the degrees of freedom that are directly related to the thermalized
CFT. The variation of $E$ with respect to $\mu$ eliminates the contribution from
the Casimir energy.

\section{Time-dependent boundary}

\subsection{Metric}
We are interested in generalizing the previous discussion to the case 
of a time-dependent boundary. If we assume that the spatial part is maximally 
symmetric, the only allowed possibility is
a boundary with the 
form of a FLRW spacetime
\be\label{eqmetricb} ds_0^2 = g_{\mu\nu}^{(0)} dx^\mu dx^\nu = 
-d\tau^2 + a^2(\tau) d\Omega_k^2 \ .\ee
In order to apply holographic renormalization, 
we choose a different foliation away from the black hole, 
consisting of hypersurfaces whose metric is asymptotically of the form (\ref{eqmetricb}).
To achieve this we need to make a change of coordinates 
$(t,r) \to (\tau, z)$ and bring the black-hole metric (\ref{eqmetric}) 
to the form
%
\be\label{eqmetric3} ds^2 = \frac{1}{z^2} \left[ dz^2 
- \mathcal{N}^2(\tau,z) d\tau^2 + \mathcal{A}^2 (\tau,z) d\Omega_k^2 \right]\ , \ee
where $\mathcal{N}, \mathcal{A} \geq 0$ and 
$\mathcal{N}(\tau,z)\to 1$, $\mathcal{A}(\tau,z)\to a(\tau)$ 
as we approach the boundary $z=0$.
Comparison with the static case (eq.~(\ref{eq4a})) suggests the ansatz
\be \mathcal{A}^2 = \alpha(\tau) + \beta(\tau) z^2 + \gamma (\tau) z^4 \ , \ee
where $\alpha(\tau), \beta(\tau), \gamma(\tau)$ are functions to be determined.
The solution of the Einstein equations gives \cite{ast}
\be \label{eqgam}
\mathcal{N} = \frac{\dot{\mathcal{A}}}{\dot{a} } \ , ~~~~~~
\alpha  = a^2  \ , ~~~~~~
\beta = - \frac{\dot a^2 + k}{2} \ , ~~~~~~
\gamma = \frac{(\dot a^2 +k)^2 +4\mu}{16a^2} \ , \ee
where we fixed the integration constants by comparing with the 
static case (eq.~(\ref{eq4x})).

The derivation of the metric (\ref{eqmetric3}) has been achieved 
without reference to the static AdS-Schwarzschild metric of eq. (\ref{eqmetric}).
The two metrics 
agree provided that 
\bes\label{eqsys1} \frac{(r')^2}{f(r)} - f(r) (t')^2 &=& z^{-2} \nonumber\\
\frac{r' \dot r}{f(r)} - f(r) t'\dot t &=& 0 \nonumber\\
\frac{\dot r^2}{f(r)} - f(r) \dot t^2 &=& - \mathcal{N}^2 z^{-2} \nonumber\\ r 
&=& \mathcal{A}z^{-1} \ . \ees
The last equation fixes $r(\tau,z)$. Two of the other three equations may 
then be used to determine the derivatives $\dot t$ and $t'$. 
We obtain
\be\label{eqsys2} 
\dot t = =-\epsilon \frac{\dot{\mathcal{A}} r'}{f \dot a} = -\epsilon \frac{{\mathcal{N}} r'}{f }\ , 
\ \ \ \ \ \ \ \ 
t' = -\epsilon \frac{\dot a}{zf} \ ,\ee
with $\epsilon=\pm 1$.
In fact, these expressions satisfy all three equations.  
Outside the event horizon, where $f(r)>0$ and $r'<0$,
It is natural to impose that $t$ and $\tau$ increase simultaneously. This requires $\epsilon=1$. 
We shall not need explicit expressions for the functions $t(\tau,z)$, $r(\tau,z)$,
because we shall base our discussion on the metric 
in the ($\tau,z$) coordinates (eqs.~(\ref{eqmetric3}) - (\ref{eqgam})). 
However, it must
be emphasized that the coordinate transformation is singular at the location of the 
event horizon, where $f(r_e)=0$.

\subsection{Horizons}

The coordinates ($\tau,z$), in terms of which the AdS-Schwazschild metric takes the
time-dependent form (\ref{eqmetric3}), are a 
generalization of the isotropic coordinates we considered
in the previous section.  
This is obvious, as eq. (\ref{eqmetric3}) reduces to
eq. (\ref{eqmetric1}) for $a=1,\dot{a}=\ddot{a}=0$. 
The difference with the static case is that now the 
coordinate $z$ spans a larger
part of the Schwarzschild geometry. From eqs. (\ref{eqsys1}) we obtain
\be \label{rprime}
\left(r' \right)^2=\frac{\dot{a}^2+f(r)}{z^2}.
\ee
For a time-dependent boundary, $\partial r/\partial z$ does not vanish at the location
of the event horizon $r_e$. Instead, for a given value of $\tau$, it vanishes at 
\be \label{zm}
z^2_m(\tau)=\frac{4a^2(\tau)}{\left(\left( \dot{a}^2+k \right)^2+4 \mu \right)^{1/2}}.
\ee
In all the cases that we have studied, this value corresponds
to some (in general time-dependent) 
$r_m$ smaller than $r_e$. As a result, the new coordinate $z$ spans a patch 
that reaches behind the event horizon.
The region $(r_m,\infty)$ of $r$ is covered twice by the
coordinate $z$ taking values in $(0,\infty)$.  Similarly to the static case discussed in section 2, for constant $\tau$ we have
a Einstein-Rosen bridge connecting two asymptotically flat regions. The difference is that the 
minimal distance from the singularity is
smaller than the horizon distance in this case. We emphasize the established fact that incoming 
photons, emanating from one of the flat regions, cannot cross the bridge and emerge in the other flat region.
They always fall into the singularity \cite{mtw}. 

A surface that will be important in the following is defined by the function
$z_a(\tau)$ that satisfies $\mathcal{N}(\tau,z_a(\tau))=0$. For $\mu\not=0$, $z_a(\tau)$
is given by the smallest root of 
\be \label{zat}
1-\frac{\ddot{a}}{2a}\, z_a^2+
\frac{\left(\dot{a}^2+k \right)\left(2a\ddot{a}- \dot{a}^2- k \right)
-4\mu}{16a^4}\, z_a^4=0,
\ee
which is 
\be \label{za}
z^2_a(\tau)=\frac{4a^2(\tau)}{a \ddot{a}+
\left(\left( \dot{a}^2-a\ddot{a}+k \right)^2+4 \mu \right)^{1/2}}.
\ee
Again, in all the cases we studied $z_a(\tau)$ corresponds to values $r_a$ smaller
than $r_e$. 

An interesting example is obtained for $a(\tau)=\lx \tau$.
It is analogous to the Bjorken flow, but with isotropic expansion. 
We find constant values 
\be \label{apr}
r_m^2=r_a^2 
= \frac{1}{2}
\left[ -\kt+\left( \kt^2+4\mu \right)^{1/2}\right],
\ee
where $\kt=k+\lx^2$.
On the other hand, the event horizon has
\be \label{evr}
r_e^2 =  \frac{1}{2}
\left[ -k+\left( k^2+4\mu \right)^{1/2} \right].
\ee
It can be checked that $r_m=r_a \leq r_e$. 
For $\lx=0$ all three surfaces defined by $r_m$, $r_a$ and $r_e$ coincide.

We have $z_a(\tau,z) < z_m(\tau,z)$ for accelerating expansion ($\ddot{a}>0)$, and
$z_a(\tau,z) > z_m(\tau,z)$ for decelerating expansion $(\ddot{a}<0)$.
For decelerating expansion the corresponding values $r_m={\mathcal{A}}(\tau,z_m)/z_m$ and 
$r_a={\mathcal{A}}(\tau,z_a)/z_a$ almost coincide. 
This can be seen explicitly for the cosmological spaces with 
$a=\tau^{\nu}$ and constant $\nu$ for large $\tau$. For example, for $\nu=1/2$ and $k=0$ the values of $r_m$ and $r_a$
differ in the third order of the late-time expansion:
\bes 
r_m&=&r_e \left(1-\frac{1}{16}\frac{1}{\sqrt{\mu}\tau}
+\frac{1}{512}\frac{1}{\mu\tau^2}\, ... \right),
\label{mpponehalf} \\
r_a&=&r_e \left(1-\frac{1}{16}\frac{1}{\sqrt{\mu}\tau}
+\frac{3}{512}\frac{1}{\mu\tau^2}\, ... \right),
\label{apponehalf} 
\ees
with $r_e=\mu^{1/4}$. 
Asymptotically, $r_m=r_a=r_e$.

We would like to determine the horizons of the metric (\ref{eqmetric3}).
The coordinates that we are employing are not the most appropriate ones for
this purpose. Their choice has been enforced by the necessity of a 
Fefferman-Graham parametrization of the metric near the boundary. 
A generalization of the 
Eddington-Finkelstein coordinates that would reduce to Fefferman-Graham ones near the
boundary would be more appropriate, but is difficult to derive. 
In the case of the Bjorken flow, the parametrization of the metric in terms of
Eddington-Finkelstein coordinates has been achieved only through a perturbative expansion
\cite{kinoshita,hubeny1,hubeny2}. It is possible though to derive the basic physics
independently of the parametrization. For example, 
the original
proposal for the gravity dual to the Bjorken flow employs isotropic coordinates, 
similar to ours \cite{janik}. 

Despite the misgivings about the choice of coordinates, we may push along and
try to identify possible horizons. 
An apparent horizon is defined as the boundary of the region of
trapped surfaces. This boundary is a surface on which 
the expansion of outgoing null geodesics vanishes.
The out/ingoing null geodesics obey
$(dz(\tau)/d\tau)_\pm=\mp\mathcal{N}(\tau,z)$ and
define a surface of areal radius $A(\tau,z(\tau))/z(\tau)=r(\tau)$. 
The growth of the 
volume of this surface is proportional to the total time derivative of $r$ along 
the light path, i.e. to 
\be \label{growth}
\left(\frac{dr}{d\tau}\right)_\pm=\dot{r}+r'\left(\frac{dz}{d\tau}\right)_\pm
=\mathcal{N}\left( \frac{\dot{a}}{z} \mp r' \right),
\ee 
with $r'$ given by eq. (\ref{rprime}).
The product 
$(dr/d\tau)_+(dr/d\tau)_-=-\mathcal{N}^2 f(r)/z^2$, which is expected to be invariant under
relabelling of the scalars that define null hypersurfaces \cite{kinoshita},  
vanishes on two surfaces. One of them is the event horizon, located at $r=r_e$, with $f(r_e)=0$.
The other is the
surface parametrized by $z_a(\tau)$, for which $\mathcal{N}=0$. 

Two remarks are important for the interpretation of eq. (\ref{growth}):
\\
a)
The total $\tau$-derivative of $t$ along a null geodesic is given by
\be \label{dtdtau}
\left(\frac{dt}{d\tau}\right)_\pm=\dot{t}+t'\left(\frac{dz}{d\tau}\right)_\pm
=\pm\epsilon\frac{\mathcal{N}}{f(r)}\left( \frac{\dot{a}}{z} \mp r' \right)=
\epsilon \left| \left(\frac{dr}{d\tau}\right)_\pm\right|\frac{1}{f(r)}.
\ee 
For $\epsilon=1$, $\tau$ and $t$ flow in the same direction on a null geodesic outside the horizon ($f(r)>0$).
On the other hand, they flow in opposite directions inside the horizon ($f(r)<0$). Of course, $t$ is not a temporal
coordinate in the second case. Using eqs. (\ref{growth}), (\ref{dtdtau}) we can see that
our definition of out/ingoing null geodesics  
translates into $(dr/dt)_\pm=\pm\epsilon f(r)$. For $\epsilon=1$, this is the standard definition in 
$(t,r)$ coordinates.
\\
b) For small $z$, we have $r'<0$. However,
for $\dot{a}>0$ we have $z_m<z_a$, so that $r'>0$ in the interval $z_m<z\leq z_a$. 
It may possible then for an incoming null ray to move from a region with $r'<0$ into a region with $r'>0$, so that  
it would cross the Einstein-Rosen bridge. It is known that this situation cannot occur \cite{mtw}. We have seen
earlier that, for decelerating expansion, $r_m$ (where $r'=0$) and $r_a$ (where $\mathcal{N}=0$) almost coincide, so that
$dr/dt$ along a null geodesic (given by eq. (\ref{growth})) approaches zero in the vicinity of these points.
Even though an explicit proof is impossible because of the complexity of the metric, there is no reason to expect that
the general conclusions about the crossing of the Einstein-Rosen bridge may violated in our setup. 
\\
The above considerations indicate that we can take $\epsilon=1$, $r'<0$ in our analysis.

The outgoing geodesics have positive expansion at the location of the
event horizon of the metric (\ref{eqmetric}), where $f(r_e)=0$ and $r'=-\dot{a}/z$. 
This surface is parametrized now by a function $z_e(\tau)$. On the other hand, the
ingoing geodesics have zero expansion there. This behavior is identical to that encountered for the standard 
four-dimensional Schwarzschild metric in outgoing Eddington-Finkelstein coordinates. The outgoing
null geodesics are well-behaved on the horizon, while the horizon is made up from world lines of
ingoing photons \cite{mtw}. It is worth noticing that the situation is reversed for contracting boundary
metrics with $\dot{a}<0$. In this case the event horizon is formed by outgoing photons, while the incoming
ones are well-behaved on it. This is identical to what happens in the  
four-dimensional Schwarzschild geometry in ingoing Eddington-Finkelstein coordinates.
Of course, there is nothing mysterious behind this behavior. It originates in the transformation 
from $(t,r)$ to $(\tau,z)$, which is singular on the horizon. It would take an infinite time $t$ for any 
geodesic to cross the horizon.

The expansion of all null geodesics also vanishes on the
surface parametrized by $z_a(\tau)$, for which $\mathcal{N}=0$. 
We have seen that the nature of this surface is strongly linked to the evolving Einstein-Rosen
bridge appearing in the system of coordinates we are employing. The fact that no information 
can be retrieved from behind this surface  means that it could define  
the boundary of trapped surfaces, and can be characterized as an apparent horizon.
The limitations imposed by our coordinate system
do not allow us to derive more concrete information on the nature of 
this speculative apparent horizon. 
We mentioned earlier 
that the surface with $\mathcal{N}=0$ is located behind the event horizon in all the models we have studied. 
This is consistent with the expectation that an apparent horizon can exist only behind an event
horizon \cite{ellis,wald}. 
The fact that the apparent horizon is located at a smaller
areal distance than $r_e$ has another implication: If the entropy in a dynamical 
situation is related to the 
area of the apparent horizon, then it is smaller than the maximal
entropy of a black hole, given by the area of the event horizon.

We emphasize at this point that we do not derive the entropy of the CFT by 
identifying it with the area of one of the horizons we discussed. Our approach 
in the following will 
be to compute the entropy directly from the energy density and the temperature.
Only then we shall examine whether the derived entropy is proportional
to a given area.

\section{Temperature}
\subsection{General expression}
A thermalized system fluctuates at the microscopic level with  
a characteristic time scale of order $1/T$, where $T$ is its temperature. For strongly
coupled theories, like the ones we are considering, this scale determines the interaction
rates that help maintain the system thermalized. At the macroscopic level, the 
system, including its temperature, may evolve with a different, much longer, 
characteristic time scale. A typical 
example is the Universe at early epochs, during which several of its energy components 
were thermalized. 
For the problem we are considering, 
a Hawking temperature $T$ can be assigned to the AdS-Schwarzschild solution with a 
time-dependent boundary in the 
limit that the variation of the scale factor $a(\tau)$ 
is negligible at time intervals of order
$1/T$. This requires $T\gg \dot{a}/a$.

The calculation of the temperature in section 2 for the static case can 
be repeated by approximating 
$a(\tau)$ and its time derivatives as constant. 
A conical singularity appears again, but now it is located 
at the position of the apparent horizon $z_a(\tau)$. 
Eliminating the
singularity in Euclidean space
around the apparent horizon for $\mu\not=0$ gives
\be \label{hts}
T(\tau)=\frac{1}{2\pi}\left|
\frac{-4a^3\ddot{a}z_a+
\left[\left(\dot{a}^2+k \right)\left(2a\ddot{a}- \dot{a}^2-k \right)
-4\mu\right]z_a^3
}{4a^3\mathcal{A}_a}
\right| 
=\frac{1}{2\pi}
\left| \frac{4a-\ddot{a}z^2_a}{\mathcal{A}_a z_a} \right|\, ,
\ee
where 
$\mathcal{A}_a=\mathcal{A}(\tau,z_a)$ and we have 
used eq. (\ref{zat}). 
For constant $a=1$ the apparent and event
horizons coincide, and eq. (\ref{hts})
reproduces the Hawking temperature in the static case, given by eq. (\ref{htz}). 
For time-dependent $a$ the evolution of the apparent horizon and
the temperature depend on the nature of the expansion. 

\subsection{Zero acceleration}
The simplest case is analogous to the Bjorken flow, but for an 
isotropic expansion. The scale factor is
$a(\tau)=\lambda \tau$ with 
$\lambda$ constant. The apparent horizon is defined by 
\be \label{apz}
z_a^2  = \frac{4a^2}{\left( \kt^2 +4 \mu \right)^{1/2}},
\ee
where $\kt=k+\lx^2$, or equivalently by eq. (\ref{apr}).

The temperature is given by 
\be \label{ttt} 
T=\frac{1}{\sqrt{2}\, \pi a} \left(
\frac{\kt^2+4\mu}{\left(\kt^2+4\mu \right)^{1/2}-\kt}
\right)^{1/2}.
\ee
Its time evolution is equivalent to a redshift by the scale factor $a(\tau)$.
However, the proportionality constant is not just the temperature in the static 
case, given by eq. (\ref{htz}). The two expressions differ by the change of
the effective curvature $k\to \kt=k+\lx^2$. This modification is natural as the
total curvature of the boundary metric is proportional to $\kt$ for $a=\lx \tau$.
For sufficiently large $\lambda$ we have $\kt >0$ for any value of $k$. 
For this reason we expect behavior similar to that of a CFT on a sphere.
A first consequence is that the temperature diverges for $\lambda^4 \gg \mu$.
This is completely analogous to the divergence of the temperature for a
static background with $k=1$ and $\mu \to 0$. 

The influence 
of the curvature resulting from the expansion on the temperature
is an important physical effect revealed by our study. The expansion does not 
result in a simple redshift of the temperature, but induces additional corrections. 
These are reflected on the entropy as well.

A comment is due at this point about the validity of our assumption $T\gg \dot{a}/a$ 
during a fast cosmological expansion. The essential requirement is that the 
rate of the interactions 
that keep the system thermalized be faster than the expansion
rate of the space-time. For a strongly coupled theory with a very large number of
fields, such as the one we are considering, the interaction rate is proportional to
$T$, with a proportionality coefficient of order 1 or larger. Unitarity 
does not allow this coefficient to grow indefinitely for $N\to \infty$, but we find 
it plausible that it is larger than 1. For $a=\lx \tau$, $k=0$, we have 
\be \label{validity}
\frac{T}{\dot{a}/{a}}=\frac{1}{\sqrt{2}\, \pi } \left(
\frac{1+4\mu/\lx^4}{\left(1+4\mu/\lx^4 \right)^{1/2}-1}
\right)^{1/2}.
\ee
Both for $\lx\ll \mu^{1/4}$ and $\lx \gg \mu^{1/4}$ we have $T/(\dot{a}/a)\gg 1$. These
two limits suffice for establishing the basic features of the entropy in the following
section. We also find it plausible that the system remains thermalized
even for comparable $\lx$ and $T$.

One word of caution is also important. For $\lx \gg \mu^{1/4}$ the temperature is computed
through a conical singularity at $r_a\sim (\mu^{1/4}/\lx)\, \mu^{1/4} \ll r_e$, within a region of high curvature.
Even though our approach gives corrections to the temperature that may be interpreted as arising from
the curvature, the quantitative accuracy of the results is not guaranteed near the black-hole singularity.
For this reason, our results concerning boundary metrics that expand very fast (e.g. at late times of 
an accelerated expansion) must be considered as qualitative.

\subsection{Non-zero acceleration}
For a general $a(\tau)$ the evolution is more complicated. 
Of particular interest are the cosmological spaces with 
$a=\tau^{\nu}$ and constant $\nu$ for large $\tau$. We concentrate on the 
case $k=0$. The cases with $k=\pm 1$ can be studied along similar lines.
For $0<\nu<1$ the expansion is decelerating and for $\tau \to \infty$ we
always have $\dot{a}^4 \ll \mu$.
When this happens, the curvature of the boundary geometry becomes 
negligible compared to the scale set by the thermal energy of the CFT. In the same limit
the apparent horizon approaches the event horizon.
For example, for $\nu=1/2$ we find that $r_a$ is given by eq. (\ref{apponehalf}).
The temperature approaches that of the static case, redshifted by
the scale factor. For $\nu=1/2$ the leading terms in the late-time expansion are
\be \label{temponehalf}
T=\frac{\mu^{1/4}}{\pi a} \left( 
1+\frac{1}{16}\frac{1}{\sqrt{\mu}\tau}+\frac{15}{512}\frac{1}{{\mu}\tau^2}\, ...
\right).
\ee

For $\nu>1$ the expansion is accelerating and at late times we have $\dot{a}^4\gg \mu$.
The apparent horizon deviates strongly from
the event horizon and $r_a$ eventually approaches zero. 
For example, for $\nu=2$ (constant
$\ddot{a}$) the leading behavior is 
$r_a=2\sqrt{\mu}/(3\tau)$. 
The product $Ta$ diverges asymptotically for  
$\tau \to \infty$. We discussed the origin of this behavior in the case $a=\lx\tau$ with
$\lambda \to \infty$: The regime $\dot{a}^4 \gg \mu$ is equivalent to the 
$\mu \to 0$ limit for the static case with $k=1$. For $\nu > 1$ the solution always
approaches this regime at late times, which leads to the asymptotic divergence of 
$T a$.
For $1<\nu<3/2$ the temperature falls with a power of the scale factor 
smaller than 1. 
For $\nu=3/2$ we have asymptotically a constant temperature $T=9/(8\pi\sqrt{\mu})$.  
Finally, for $\nu>3/2$ the temperature grows with time and diverges asymptotically.

We observe that, apart from the rescaling by $a$,
there are two qualitatively different types of evolution. 
For $\nu <1$
the CFT corresponds to a black hole with a mass that grows relative to the
scale of the curvature induced by the expansion. 
For $\nu>1$ the effective mass of the black hole seems to
diminish and eventually vanish for $\tau \to \infty$. 
It will be clearer in the following that the two quantities that characterize the
different types of evolution are the Casimir and the thermal energy of the CFT.
For $\nu<1$ the Casimir energy becomes negligible at late times, while for 
$\nu>1$ it dominates over the thermal energy.

One should keep in mind
that the small-mass AdS-Schwazschild solution in the static case is unstable \cite{page}. 
The large-mass solution with the
same temperature is dual to the deconfined phase of the CFT. 
It seems reasonable to interpret the 
black-hole configuration with an accelerating boundary as dual to
a CFT in the deconfined phase on an accelerating FLRW background geometry.
The fact that for such a configuration the black-hole mass diminishes with time relative to the curvature makes it likely that 
the corresponding CFTcannot exist in a stable phase. 
In the following we shall find more indications of this problematic behavior.

\subsection{dS boundary} 
The case $a= \exp(H\tau)$, $k=0$ corresponds to a de Sitter (dS) boundary. 
For $\mu\not=0$ and large 
$\tau$ we find that the temperature quickly approaches the value  
$T=H/(\sqrt{2}\, \pi)$, which differs from the standard temperature of dS space by
a factor $\sqrt{2}$. This means that, even though $\mu/a^4$ becomes negligible relative to $(\dot{a}/a)^4=H^4$,
the temperature of the CFT does not approach the dS temperature. 
The discrepancy indicates that the configuration with $\mu\not= 0$ 
on a background with $a= \exp(H\tau)$ cannot
evolve continuously to pure dS space. This conclusion is consistent with our earlier argument that
a deconfined CFT on an accelerating background is not in a stable phase. 

A crucial observation is that the limits $\mu\to 0$ and $\tau \to \infty$ are not equivalent for a dS boundary.
Further insight can be obtained if 
we set $a= \exp(H\tau)$, $k=0$, $\mu=0$ directly in the metric (\ref{eqmetric3}).
In this way we find 
$\mathcal{N}(\tau,z)=1-{H^2}z^2/{4}$. 
For the determination of the temperature we can follow the same procedure as before.
Despite the absence of
a black hole, a conical singularity still exists at
$z_{a}=2/H$ for periodic Euclidean time. 
The location of the singularity is $\tau$-independent
for a dS boundary. As a result, we need not make any assumptions about the 
relative size of $T$ and $H$. 
The singularity can be eliminated for an appropriate value of the temperature.
We find $T=H/(2\pi)$. 
It is
remarkable that the correct temperature of four-dimensional dS space can be reproduced 
through five-dimensional AdS space, when the coordinate frame defined by
eq. (\ref{eqmetric3}) is used.

\section{Entropy}
\subsection{Stress-energy tensor and entropy}
The stress-energy tensor of the dual CFT on the cosmological boundary (\ref{eqmetricb}) 
is determined via holographic renormalization (eq.~(\ref{eq3a})). 
We obtain the energy density and pressure \cite{ast}, respectively, 
\bes \langle ( T^{(CFT)} )_{\tau \tau} \rangle 
&=& \frac{3}{64\pi G_5} \ \frac{(\dot a^2+k)^2 + 4\mu}{a^4} 
\label{eq4aa} \\ 
\label{eq4bb} 
\langle ( T^{(CFT)} )_{i}^i \rangle &=&  \frac{(\dot a^2 + k)^2+4\mu 
-4a\ddot a (\dot a^2 + k)}{64\pi G_5a^4} \ ,\ees
where no summation over $i$ is implied.
We deduce the expected conformal anomaly
\be g^{(0)\mu\nu} \langle T_{\mu\nu}^{(CFT)} \rangle = 
- \frac{3\ddot a (\dot a^2 + k)}{16\pi G_5a^3} \ .\ee

The boundary geometry can be made dynamical if one introduces an Einstein term
for the boundary metric, with a four-dimensional Newton's constant $G_4$, 
and employs mixed boundary conditions \cite{marolf}.
The resulting Friedmann equation is \cite{ast,kiritsis}
\be \label{friedmann}
\left(\frac{\dot{a}}{a} \right)^2 + \frac{k}{a^2} 
= \frac{8\pi G_4}{3}
\left\{ \frac{1}{16\pi G_5} \left[ \frac{\left( \dot{a}^2+k \right)^2}{a^4} 
+ \frac{4\mu}{a^4} \right] +  \rho \right\}\ .\ee
The Casimir energy density is $\sim(\dot{a}^2+k)^2/a^4$ and
reflects the total curvature of the boundary metric. For $\dot{a}^4 \gta \mu$
it becomes comparable to or dominates over the
thermal energy $\sim \mu/a^4$ of the CFT. The energy density $\rho$ corresponds to
all energy components that are located on the boundary. 

In this work we do not consider a dynamical boundary metric. Instead we take the 
form of $a(\tau)$ as an input and investigate the properties of the CFT on such
a background. For a dynamical boundary the energy density $\rho$ must
be the dominant component, with an appropriate equation of state that will result in
the desired form of $a(\tau)$. This means that the dominant contribution to the 
entropy of the system will come from the boundary energy component. 

We would like to derive the entropy of the CFT in analogy to the static case. 
For this we follow a generalization of the intuitive approach of section 2.
The entropy now depends on the parameter $\mu$ and the time $\tau$ through the 
scale factor $a(\tau)$ and its derivatives. We consider an infinitesimal adiabatic
variation
of $\mu$ that takes place within a time interval that is sufficiently small for
the evolution of $a(\tau)$ to be negligible. In contrast to the determination of the 
temperature, the required time for the variation can be made arbitrarily small by
sending $d\mu \to 0$. 
The fundamental relation $dE+pdV = TdS$ can be employed for the determination of
the entropy. The volume $a^3 V_k$ of the boundary remains constant, while 
the temperature is given by eq. (\ref{hts}).
The location of the apparent horizon satisfies 
\be \label{dzdmu}
dz_a=-\frac{z_a^5}{4a^3(4a-\ddot{a}z^2_a)}d\mu=-\frac{z^4_a}{8\pi a^3  \mathcal{A}_a}
\frac{d\mu}{T}=-\frac{2G_5}{3V_k a^2}\frac{z_a^4}{\mathcal{A}_a}\, dS,
\ee
where we have employed eqs. (\ref{zat}), (\ref{hts}), (\ref{eq4aa}) and $dE/T=dS$.
Making use of eq. (\ref{zat}), 
we can express $\mathcal{A}_a$ as a function of $z_a$ only, with implicit dependence
on $\mu$. It is
given by
\be \label{Aaa}
\mathcal{A}_a(z_a)=\left(
2a^2-\frac{\dot{a}^2+k+a\ddot{a}}{2} z^2_a 
+\frac{(\dot{a}^2+k)\ddot{a}}{8a}\, z^4_a
\right)^{1/2}.
\ee
We can now integrate eq. (\ref{dzdmu}) in order to determine the CFT
entropy $S$ as a function of the location of the apparent horizon $z_a$.
We obtain
\be \label{solen}
S=\frac{V_k}{4\,G_5}\left(\frac{\mathcal{A}_a}{z_a}\right)^3
-\frac{3V_k}{32\,G_5}\frac{\left(\dot{a}^2+k \right)\ddot{a}}{a}
\int^{z_a} \mathcal{A}_a(z) dz +F(a,\dot{a},\ddot{a}),
\ee
with $A_a(z)$ defined in eq. (\ref{Aaa}). The expression can be checked through
simple differentiation. We emphasize again that $\tau$ is kept 
constant during these manipulations.

The constant of integration in the above expression is an  
arbitrary function of $a(\tau)$ and its derivatives. It appears because we have 
determined the CFT entropy by considering its variation with respect to $\mu$. 
The form of $F(a,\dot{a},\ddot{a})$ can be fixed if we define the value of the entropy for
a specific value of $\mu$. We postulate that the CFT entropy vanishes for $\mu\to 0$. This is consistent
with the fact that the thermal energy of the CFT (after the Casimir energy has been subtracted from
$\langle ( T^{(CFT)} )_{\tau \tau} \rangle$ in eq. (\ref{eq4aa})) vanishes in this limit.

We emphasize that the boundary geometry is fixed in our calculation, as we employ the
standard AdS/CFT correspondence. Within the limitations of such an approach we
can compute only the entropy of the CFT. The entropy of the component that would drive
the assumed expansion in a dynamical setting cannot be computed.

The first term in eq. (\ref{solen}) describes a contribution that 
is proportional to the area of the apparent horizon. This 
is time dependent for a general cosmological background. 
The integral in the second term
can be expressed in terms of incomplete elliptic integrals. Unfortunately,
the resulting expression is not very illuminating. For this reason, it is
more instructive to consider specific cases, as we did for the temperature.

\subsection{Specific cases}

The expression for the entropy becomes
transparent for $a(\tau)=\lambda \tau$. In this case we have $\ddot{a}=0$, and
eq. (\ref{solen}) becomes 
\be \label{entra}
S=\frac{V_k}{4G_5}\left(\frac{\mathcal{A}_a}{z_a} \right)^3
=\frac{V_k}{4G_5}r_a^3,
\ee
where the areal distance of the apparent horizon $r_a$ is constant, 
given by eq. (\ref{apr}).
We have set $F=0$ in order to
obtain consistency with eq. (\ref{entropy}) for $\lambda=0$.
The total 
entropy is proportional to the constant area of the apparent horizon. 
The entropy density
$s=S/(a^3 V_k)$ scales as $a^{-3}$.
This particular example is the one closest to the Bjorken flow. The 
scale factor is linear in time, even though the expansion is isotropic. We 
find strong similarity with the conclusions of \cite{hubeny2}. 
There is an apparent horizon that is distinct from the event horizon.
It is the former that is relevant for the definition of the entropy of the
dual CFT. In our example we were able to calculate the entropy, using
the stress-energy tensor and the value of the temperature that eliminates the conical 
singularity of the five-dimensional geometry.

A class of boundary geometries with relevance for 
cosmology are the spaces with 
$a=\tau^{\nu}$ and constant $\nu$ for large $\tau$. We concentrate on the 
case $k=0$. The cases with $k=\pm 1$ can be treated along similar lines.
In order to obtain the $\mu$-dependent part of the entropy we differentiate 
eq. (\ref{solen}) with respect to $\mu$, expand in powers of $1/\tau$ and
integrate with $\mu$ again. 
The resulting entropy displays different behavior depending on whether $\nu<1$ or 
$\nu>1$. We present various examples that display the main characteristics. 

For $\nu<1$ the expansion is decelerating ($\ddot{a}<0$).
A typical example has $\nu =1/2$. For a dynamical boundary, it would
correspond to a homogeneous and isotropic space-time whose expansion is 
driven by relativistic matter, such as the thermalized CFT. In such a case no 
new matter component needs to be introduced on the boundary to drive the
expansion. 
For the CFT entropy on such a background
we obtain
\be \label{solhalf}
S=\frac{V_k}{4G_5}\left[
\mu^{3/4}\left(1-\frac{3}{16}\frac{1}{\sqrt{\mu}\, \tau}
+\frac{15}{512}\frac{1}{\mu \tau^2} \, ... \right)
+\frac{3}{64}\frac{1}{\mu^{1/4}\,\tau^2}
\left(1+
\frac{1}{16} \frac{1}{\sqrt{\mu}\, \tau}-\frac{29}{2560}\frac{1}{\mu \tau^2}\, ... \right) 
\right],
\ee
with the two terms originating in the first and second term of eq. (\ref{solen}),
respectively. As the entropy is a non-analytic function of $\mu$ near zero, the above
series converges only for non-zero $\mu$ and large $\tau$.
The main contribution to the entropy comes from the first term in eq. (\ref{solen}). 
The second term contributes starting only at the third order of the expansion
in $1/(\sqrt{\mu}\tau)$. The first order in this expansion gives the area
of the event horizon. The second order includes a negative term that reduces the entropy 
to the area of the apparent horizon. 

The leading behavior for all decelerating expansions with 
$0<\nu < 1$ is similar to the example we described: 
The total entropy is proportional to the area of the apparent horizon. 
It grows with time and asymptotically
approaches its maximum value, proportional to the area of the event horizon. 
The entropy density scales with an additional factor $a^{-3}$ relative to the
total entropy. 
We have seen in the previous section that the temperature has similar behavior: It receives 
corrections proportional to the curvature of the boundary at early times. These become
negligible at late times and    
the temperature becomes equal to the temperature of the static case, given
by eq. (\ref{htz}), redshifted by a factor of $a$. 

For $\nu > 1$ the expansion is accelerating ($\ddot{a} >0$).
The total entropy falls off with time and asymptotically
approaches zero for $\tau\to \infty$. We give two examples displaying this behavior.
For $\nu=3/2$ we find
\be \label{solthreehalf}
S=\frac{V_k}{4G_5}\left[
\frac{8}{27}\frac{\mu^{3/2}}{\tau^{3/2}}
\left(1-\frac{2}{9}\frac{\mu}{\tau^2}
-\frac{14}{243}\frac{\mu^2}{\tau^{4}} \, ... \right)
+\frac{4}{27} \frac{\mu^{3/2}}{\tau^{3/2}}
\left(1-\frac{58}{45}\frac{\mu}{\tau^2}+\frac{2878}{1701}\frac{\mu^2}{\tau^4}\, ... \right) 
\right].
\ee
For $\nu=4$ we have
\be \label{solfour}
S=\frac{V_k}{4G_5}\left[
\frac{1}{64}\frac{\mu^{3/2}}{\tau^9}\left(1+\frac{3}{64}\frac{\mu}{\tau^{12}}
-\frac{81}{8192}\frac{\mu^2}{\tau^{24}} \, ... \right)
+\frac{3}{64} \frac{\mu^{3/2}}{\tau^9}
\left(1-\frac{33}{320}\frac{\mu}{\tau^{12}}+\frac{111}{8192}\frac{\mu^2}{\tau^{24}}\, ... \right) 
\right].
\ee
The first two terms in eq. (\ref{solen}) give
comparable contributions. To leading order, the  
entropy is again proportional to the apparent
horizon, even though the proportionality constant depends on $\nu$. For an accelerating
boundary, the areal distance of the 
apparent horizon asymptotically shrinks to zero for large $\tau$.

This behavior is consistent with the evolution of the temperature that we 
discussed earlier. For $\nu <1$, 
the CFT corresponds to a black hole with a growing effective mass that
asymptotically reaches a constant value equal to $\mu$. The entropy also becomes
constant asymptotically.
For $\nu>1$, the effective mass of the black hole 
decreases and vanishes asymptotically, along with its entropy.

For the interpretation of the above results, one should keep in mind
that the small-mass AdS-Schwazschild solution in the static case is unstable \cite{page}. 
This fact and the conclusion that the total entropy decreases with time
indicate strongly that there is a pathology in the case of the CFT on
an accelerating background. On a static background,
the large-mass AdS black hole is dual to the deconfined phase of the CFT \cite{witten}. 
It seems reasonable to 
speculate that the configuration with decreasing entropy on an 
accelerating background is again a thermalized 
CFT in the deconfined phase. Our findings indicate that this phase is probably unphysical and 
cannot exist in quasi-thermal equilibrium.
A possible interpretation is that it is unstable towards the
confined phase on the same background. 

The above explain why taking the limit
$\mu\to 0$ does not reproduce the correct temperature in the case of a
dS boundary (subsection 4.4). The state corresponding to $\mu\not= 0$ in this 
background is not connected continuously to the pure dS space.

It would be very interesting to calculate the intrinsic entropy of
the dS space-time along similar lines.
However, as we have mentioned repeatedly, 
the boundary geometry is introduced by hand in the standard holographic approach
that we have followed. This means that we have no dynamical information
on the energy density and pressure that generate the boundary geometry. 
Even though the dS temperature can be computed from the five-dimensional
geometry, we do not have sufficient information for the calculation of the entropy.

\section{Conclusions}

We have presented a calculation of the entropy of CFT on a FLRW background. 
Instead of conjecturing that the entropy is proportional to the
area of a horizon of the relevant geometry, we computed it using the energy density
and temperature. The basic framework is provided by the AdS/CFT correspondence
\cite{adscft}. The relevant gravitational configuration is the five-dimensional 
AdS-Schwazschild solution, expressed in appropriate Fefferman-Graham 
coordinates, so that the
boundary metric is of the FLRW type with a scale factor $a(\tau)$ \cite{ast}. 
The stress-energy tensor can
be computed in a straightforward manner. It reflects correctly the Casimir energy
and the conformal anomaly on this background.
The Casimir energy is proportional to the square of the total curvature of the FLRW metric
$\sim (\dot{a}^2+k)^2$. If the boundary metric is made dynamical, 
the induced corrections correspond to the ones that would result from higher-derivative
corrections to the gravitational action. This is apparent from eq. (\ref{friedmann}).

The CFT temperature can be identified with
the temperature of the black hole, as measured by an observer located near the boundary.
The Schwarzschild radial coordinate $r$ is $r=a(\tau)/z$ for a small Fefferman-Graham
coordinate $z$. 
The five-velocity of an observer located at a fixed small distance $z$ from the
boundary is $(z/a,\dot{a},\vec{0})$ in Schwarzschild coordinates. 
The boundary observer moves away
from the black hole for an expanding FLRW metric. It is natural that 
he/she would see a temperature
redshifted by the scale factor $a$ that determines the velocity. Moreover, if the expansion
velocity $\dot{a}$ is not constant, further corrections to the measured temperature
are expected because of the Unruh effect \cite{unruh}. 

Our analysis realizes all the above expectations and permits the determination of the 
corrections to the temperature. The black-hole temperature, even
in the static case, has a dependence on the curvature $\sim k$ of the boundary 
(which is proportional to that of the horizon for a given slicing). In the
time-dependent situation, the total curvature of the boundary $\sim (\dot{a}^2+k)$
becomes the relevant quantity for the determination of the temperature. 
It plays a role whenever the Casimir energy gives a significant contribution to the total
CFT energy. It is remarkable that this definition of the temperature can reproduce 
correctly the temperature of dS space, if we employ an appropriate slicing of
the five-dimensional AdS space and assume that the black-hole mass vanishes.

When the corrections to the temperature are taken into account, the total entropy
cannot be identified with the area of the event horizon. Instead, a new surface appears,
on which the expansion of outgoing null
geodesics vanishes.
The Fefferman-Graham
coordinates are not appropriate for an unambiguous determination of the true nature
of this surface. 
The interpretation we suggested
is that it represents a local apparent horizon.
This is consistent with what has been found
for the Bjorken-flow geometry \cite{kinoshita,hubeny1,hubeny2}.

The entropy of the CFT can be computed starting from the temperature and the
energy density of the CFT, which are known. We found three distinct types of
behavior. 
\begin{itemize}
\item
For a decelerating expansion of the FLRW background
($\ddot{a} < 0$) the total entropy at late times is proportional to the
area of the apparent horizon, with the standard proportionality constant.
The entropy is always smaller than the maximal entropy of the black hole, which is
proportional to the area of the event horizon. The reason is that the apparent 
horizon is always located behind the event horizon, and therefore has a smaller
area. As the rate of expansion decreases, the apparent horizon approaches the
event horizon, while its area increases. Asymptotically, the two become
identical. In this limit, the temperature and entropy density scale with
simple powers of the scale factor $a$. 
\item 
For an accelerating expansion ($\ddot{a}>0$) the  CFT entropy
decreases with time. 
This is a sign that the corresponding
configuration is unphysical. Another indication is that 
the mass of the black hole relative to the boundary curvature 
decreases. Small-mass AdS black holes are known to be unstable \cite{page}.
One possible interpretation is that this configuration corresponds to a high-temperature
CFT in the deconfined phase on an accelerating background. One may speculate that
this phase cannot be realized
in quasi-thermal equilibrium on an accelerating background, because it is 
unstable, probably towards the confined phase on the same background. 
\item 
In the case with zero acceleration ($\ddot{a}=0$) the boundary
curvature is proportional to a constant factor $\sim (\dot{a}^2+k)$.
This results in a constant entropy, which is 
smaller than the area of the event horizon. In completely analogous fashion,
the temperature has the simple scaling $\sim 1/a$, but the proportionality constant
is different from the temperature of the static case. This case is the
isotropic analogue of the Bjorken-flow geometry. 
\end{itemize}

Our results are consistent with the notion that the apparent horizon is the
relevant entity for the determination of the entropy in time-dependent situations.
The teleological character of the event horizon makes it appropriate only for 
static situations, in which there is no macroscopic time evolution. An unsatisfactory
element in our analysis is that the Fefferman-Graham system of coordinates that we
employed does not allow a thorough investigation of the properties of the apparent
horizon. A reformulation in a coordinate system more appropriate for this purpose
seems difficult, but would be very useful for the completion of the global picture.

\section*{Acknowledgments}
I would like to thank P. Apostolopoulos, V. Hubeny, 
S. Shenker, G. Siopsis, L. Susskind, S. Trivedi 
for useful discussions. 
N.~T. was supported in part by the EU Marie Curie Network ``UniverseNet'' 
(HPRN--CT--2006--035863).

\end{document}